\documentclass[aps,nofootinbib,amsmath,prd,twocolumn,showpacs,superscriptaddress,groupedaddress]{revtex4-1}

\usepackage[american]{babel}
\usepackage{amsfonts}
\usepackage{amsmath}
\usepackage{amssymb}
\usepackage{epsfig}
\usepackage{latexsym}
\usepackage{paralist}
\usepackage{fancyhdr}
\usepackage{graphicx}
\usepackage[vcentermath]{youngtab}
\usepackage{young}
\usepackage{ytableau}
\usepackage{etex}
\usepackage{braket}
\usepackage{float}
\usepackage{slashed}
\usepackage{xcolor}
\usepackage{soul}
\usepackage{dcolumn} 
\usepackage{physics}
\usepackage{bm}
\usepackage{nicefrac}

\newcommand{\comment}[1]{ }

\def\bea{\begin{eqnarray}} 
\def\eea{\end{eqnarray}}
\def\be{\begin{equation}} 
\def\ee{\end{equation}} 
\def\ba{\begin{array}}
\def\ea{\end{array}}

\def\be{\begin{equation}}
\def\ee{\end{equation}}
\def\bea{\begin{eqnarray}}
\def\eea{\end{eqnarray}}

\usepackage{amsmath}

\begin{document}

\title{A different kind of four-dimensional brane for string theory}

\author{Matteo Romoli}
\email{matteo.romoli@uniroma3.it,}
\affiliation{
Universit\`a degli Studi Roma Tre and INFN - Sezione di Roma Tre, Via della Vasca Navale 84, 00146 Roma, Italy}

\author{Omar Zanusso}
\email{omar.zanusso@unipi.it}
\affiliation{
Universit\`a di Pisa and INFN - Sezione di Pisa, Largo Bruno Pontecorvo 3, 56127 Pisa, Italy}

\begin{abstract}
%
We present a generalization of the string's Polyakov action
that describes a conformally invariant four-dimensional brane.
The new extended object is very different from the traditional D-branes
of string theory, but, nevertheless,
shares some structural similarities with the string,
especially when it comes to the low-energy limit of small tension.
We introduce a rather rich structure of tensors
that can play a role at low energies.
In analogy with the bosonic string,
we initiate the quantization of the new brane discussing
the extent in which it produces a critical dimension of spacetime
and Einstein's equations coupled to a scalar dilaton under some approximations.
\end{abstract}

\maketitle

\renewcommand{\thefootnote}{\arabic{footnote}}
\setcounter{footnote}{0}

\section{Introduction}\label{sect:introduction}

String theory has seen an enormous development over the past forty years,
much of which was achieved through two ``revolutions,'' the second of which
is still ongoing.
The theoretical foundation of string theory is based on the fact that
in the low-energy limit of the string, which is obtained as the limit
of small string tension $\alpha'$, a graviton-like mode is predicted
for the embedding spacetime, among other things \cite{Callan:1985ia}.
With opportune tweaking
of boundary conditions and symmetries,
it is eventually possible to argue that all fundamental particles,
including the graviton, could emerge as different excitations
of an otherwise unique object, which would be the string itself,
as opposed to more traditional point particle descriptions.

In these regards, several other higher dimensional objects
have been introduced over the years to support the construction of a string theory
which produces the correct degrees of freedom \cite{Dai:1989ua}.
The most important ones are probably D$p$-branes,
which are $p$-dimensional objects propagating in spacetime
described by a Nambu-Goto-like action,
and which are used to enforce the correct boundary conditions
for open strings. D$p$-branes generalize strings to the extent
in which the objects use similar geometrical actions to begin with,
but do not lead to the same predictions for the low-energy limit,
simply because D$p$-branes are not conformally invariant.
As such, branes have acquired a more auxiliary role in the construction
of string theory.

In this paper we discuss an alternative generalization of the string
that emanates directly from the Polyakov's string description \cite{Polyakov:1981rd}.
The new object is a four-dimensional brane embedded in spacetime
which enjoys a conformally invariant action built from the same degrees of freedom
of the string, that is, from its embedding function.
Classically, the new action is the unique conformally invariant
generalization of the Polyakov action with four derivatives
and, to an extent that we will clarify, it has been previously studied as a nonlinear sigma model in four dimensions
\cite{Buchbinder:1988ei,Buchbinder:1991jw,Hasenfratz:1988rf,Percacci:2009fh}.
In anticipation, we show it here in its simplest form,
which happens to be the one for a \emph{boundaryless} four-dimensional brane
\begin{equation}\label{eq:polyakov4d-intro}
\begin{split}
 S_{\rm bl} =&\frac{1}{(4\pi)^2 \alpha'} \int d^{4} x \sqrt{g}
 \Bigl\{ G_{\mu\nu} \square X^{\mu} \square X^{\nu} 
 \\&
 -2 \Bigl[ {R}^{\alpha\beta} -\frac{1}{3} R \,g^{\alpha\beta}\Bigr] G_{\mu\nu} \partial_{\alpha}X^\mu\partial_{\beta}X^\nu  
 \\&
 + \Phi_{c}(X) \,C^2
 +T_{\mu\nu\rho\theta} \partial_{\alpha} X^{\mu} \partial^{\alpha} X^{\nu} \partial_{\beta} X^{\rho} \partial^{\beta} X^{\theta} 
 \Bigr\}\,,
\end{split}
\end{equation}
where the most important ingredients are
$G_{\mu\nu}$, that is the metric of spacetime,
a scalar dilaton-like field $\Phi_{c}$, a tensor $T_{\mu\nu\rho\theta}$
with four indices, and the brane's embedding $X^\mu$,
besides metric, connection and curvature tensors on the brane itself,
whose meaning will be clarified later.
The tension of the brane is controlled by the dimensionful parameter $\alpha'$,
which has a very similar role to the analog parameter appearing in the string's action.

As a field theory, the action \eqref{eq:polyakov4d-intro}
can be interpreted as a generalized higher derivative
nonlinear sigma model in a curved four-dimensional space,
which is the four brane itself \cite{Buchbinder:1988ei,Buchbinder:1991jw}.
Quantum mechanically, the covariant approach to the quantization
of \eqref{eq:polyakov4d-intro} requires the construction of a path-integral
over the embedding degrees of freedom $X^\mu$
as well as the integration over the nonconformal degrees of freedom
of the brane's four-dimensional metric $g_{\mu\nu}$,
in the assumption that the conformal mode of the brane is
not dynamical even after quantization like for the string
\cite{Polyakov:1987zb}. This requirement hinges
on the absence of a conformal anomaly, as we shall discuss later in the paper.

In the rest of the paper we outline the main steps toward the
construction of \eqref{eq:polyakov4d-intro}, making the analogy
with the Polyakov's string as manifest as possible. For this reason,
we begin by very briefly reviewing some pivotal steps in the quantization
of the bosonic string first,
which will be recalled when discussing the four brane later on.
The literature on string theory is rather vast,
so we do not attempt to capture it in its entirety.
Furthermore, our analogy between the brane and the string holds, for the moment,
at the level of the bosonic string with critical dimension $D_{\rm cr}=26$,
which is why we limit ourselves to such model in the recap.
We will conclude by outlining some first steps toward the quantization of
\eqref{eq:polyakov4d-intro} and the difficulties that we must circumvent,
discussing also some of its possible extensions.

\section{Recap of the Polyakov action}\label{sect:2dpa}

The traditional construction of the string of string theory is based on the geometrical Nambu-Goto action.
Upon introducing an auxiliary metric on the Nambu-Goto string,
it is possible to replace its action with the classically
equivalent string's Polyakov
action, which is more suited for modern quantization purposes. We present
here some of the steps associated to the covariant analysis
of Polyakov's bosonic string
in their most basic form, which is what we need for the comparison
with the same steps undertaken in the next sections for the four-dimensional membrane. Our conventions are that Greek letters from the beginning of the alphabet
(e.g.\ $\alpha,\beta,\gamma\cdots$)
are associated to the string's coordinates (and later to the brane's),
while those at the end of the alphabet (e.g.\ $\mu,\nu,\rho\cdots$)
are associated to spacetime.

For a string embedded with coordinates $X^\mu$
in a $D$-dimensional target spacetime equipped with a metric $G_{\mu\nu}$
the Polyakov action takes the form
\begin{equation} \label{eq:polyakov2d}
\begin{split}
 S &=
 \frac{1}{4\pi\alpha'} \int d^{2} x \sqrt{g}\,
 G_{\mu\nu} \partial_\alpha X^{\mu} \partial^\alpha X^{\nu} \,,
\end{split}
\end{equation}
where we introduced coordinates $x^\alpha$ and the metric $g_{\alpha\beta}$
on the string. The action \eqref{eq:polyakov2d} is invariant under reparametrizations
if the metric $G_{\mu\nu}$ transforms as a tensor, however the true symmetries of
the Polyakov action are the isometries of the embedding that leave $G_{\mu\nu}$
invariant. We omit the famous antisymmetric Kalb-Ramond
field from \eqref{eq:polyakov2d}, though some comments on it
will come toward the end of the paper.
In the case in which the embedding is flat Minkowski's space,
then $G_{\mu\nu}=\eta_{\mu\nu}$ in some coordinates, and consequently \eqref{eq:polyakov2d}
is invariant under Lorentz transformations, besides reparametrizations of the string's coordinates. In general conserved currents
are associated with isometries of $G_{\mu\nu}$.

An important consequence of adopting the Polyakov action has to do with
the status of conformal symmetry. We have that \eqref{eq:polyakov2d} is invariant
under local Weyl transformations of the string's metric,
$g_{\alpha\beta} \to e^{2\sigma } g_{\alpha\beta}$, which is a manifestation of the fact
that the conformal degree of freedom of $g_{\alpha\beta}$ was not present
in the original Nambu-Goto action. Given that two dimensional metrics are all locally
conformally equivalent, the Weyl mode is expected to decouple from the path integral
resulting in a summation over distinct topologies \cite{Fradkin:1984pq,Fradkin:1985ys}.
Quantum conformal symmetry, however, is anomalous for \eqref{eq:polyakov2d}.
One strategy to restore it involves the generalization
of \eqref{eq:polyakov2d} to the effective action
\begin{equation} \label{eq:polyakov2d-eff}
\begin{split}
 S_{\rm eff} &=
 \frac{1}{4\pi\alpha'} \int d^{2} x \sqrt{g}\Bigl\{
 G_{\mu\nu} \partial_\alpha X^{\mu} \partial^\alpha X^{\nu} 
 + \alpha' \Phi R
 \Bigr\}\,,
\end{split}
\end{equation}
which is classically \emph{not} invariant because of the newly
introduced dilaton $\Phi(X)$,
but can be quantum mechanically invariant
if the expectation value of the trace of the stress energy tensor
on the string vanishes \cite{Callan:1985ia}. The anomaly has the form
\begin{equation} \label{eq:trace2d}
\begin{split}
 2\pi \langle \Theta^\mu{}_\mu \rangle &=
 B^{(G)}_{\mu\nu} \partial_\alpha X^{\mu} \partial^\alpha X^{\nu}
 + B^{(\Phi)} R
 \,,
\end{split}
\end{equation}
and its coefficients contain both classical and quantum contributions in the form
of the beta functions of the renormalized tensor couplings of
\eqref{eq:polyakov2d-eff},
which is understood as a generalized nonlinear sigma model
\cite{Tseytlin:1986ws}
(for an excellent recent review see \cite{Bonezzi:2021mub}).
The cancellation requires $B^{(G)}_{\mu\nu}=B^{(\Phi)}=0$.
Very roughly speaking,
from the condition $B^{(G)}_{\mu\nu}=0$ one can derive Einstein's equations
for the spacetime metric $G_{\mu\nu}$ coupled to the dilaton,
while from $B^{(\Phi)}=0$ one derives the critical dimension of the embedding as well as equations of motion for $\Phi$ itself. In particular,
\begin{equation} \label{eq:dilaton2d-coeff}
\begin{split}
 B^{(\Phi)} = \frac{D-26}{6}+{\cal O}(\alpha')
 \,,
\end{split}
\end{equation}
which has a contribution proportional to $D$ from the coordinate fields $X^\mu$
as well as a $-26$ coming from the integration of the ghosts of the nonconformal
modes of the metric $g_{\mu\nu}$.

As already stressed, the cancellation thus requires $D_{\rm cr}=26$,
which is the well-known critical dimension of the bosonic string,
while further orders in $\alpha'$ are set to zero as equations
of motion for the spacetime fields. A renowned aspect of this cancellation
is that the low energy equations of motion
can be derived from an effective action, which has the form
\begin{equation} \label{eq:string-low-energy}
\begin{split}
 S &= \frac{1}{\kappa_0}\int d^D X \sqrt{G} e^{-2 \Phi}
 \Bigl\{{\cal R}+4 G^{\mu\nu}\partial_\mu\Phi\partial_\nu\Phi
 \Bigr\} + {\cal O}(\alpha')
 \,,
\end{split}
\end{equation}
where $\kappa_0$ is a scale related to $\alpha'$ on dimensional grounds,
and further corrections are determined by higher loop computations
of the beta functions of the covariant bosonic string.
One important aspect that connects \eqref{eq:string-low-energy}
with real-world general relativity is the Curci-Paffuti relation \cite{Curci:1986hi},
which shows that the dilaton is actually constant.
Therefore we can use the Einstein frame metric,
defined $\tilde{G}_{\mu\nu}=e^{-\frac{\Phi}{D-2}} G_{\mu\nu}$,
and effectively determine the relation among the physical Planck scale,
the string's scale $\alpha'$, and the dilaton's vev.

It is worth stressing few facts before going forward.
The fields $X^\mu$ of \eqref{eq:polyakov2d}
contribute as a number $D$ of free conformal fields to the leading
part of the anomaly \eqref{eq:dilaton2d-coeff}, because in flat spacetime
the kernel of their propagation
is the Laplacian $\Delta_2 =- g^{\alpha\beta}\nabla_\alpha \partial_\beta$,
constructed with the metric $g_{\alpha\beta}$
and the compatible connection $\nabla_\alpha=\partial_\alpha+\gamma_\alpha$
on the string,
which is a Weyl-covariant operator in two dimensions. Furthermore,
with the exception of the dependencies of the spacetime tensors, the $X^\mu$
explicitly enter the actions \eqref{eq:polyakov2d} and \eqref{eq:polyakov2d-eff}
only through their derivatives $\partial_\alpha X^\mu$,
as one would expect from the action of an embedded object.
The above two facts are actually intimately connected,
because in general $d\neq 2$ the differential operator $\Delta_2$
is generalized by the Yamabe operator, which contains
a nonminimal interaction with
the curvature scalar of the form $\frac{d-2}{4(d-1)}R$.
Therefore, a conformal action in $d\neq 2$ would necessitate nonderivative
interactions for $X^\mu$,
which obviously goes against the interpretation
of $X^\mu$ as embedding coordinates.
This points to the fact that the construction of an object of different dimensionality, such as the brane that we are going to introduce soon,
might need a radical change of \eqref{eq:polyakov2d}.

\section{Pull-back and conformality}\label{sect:conformal}

As a first step toward the generalization of \eqref{eq:polyakov2d} we promote
the embedding $X^\mu$ to be the coordinates
of a \emph{four} dimensional brane. The reasons for this generalization
should become clear soon during the development of this section.
We transfer to the brane all the geometrical ingredients previously
defined for the string (e.g.\ the metric $g_{\mu\nu}$, compatible connection,
curvatures).
We also want to develop some additional geometrical ingredients
which are particularly useful when working with an embedded brane
and mutiple metrics.
Given a spacetime vector field $v^\mu$,
we can construct a new connection
\begin{equation} \label{eq:pullback-connection}
\begin{split}
 D_\alpha v^\mu &=
 \nabla_\alpha v^\mu + \partial_\alpha X^\nu \Gamma_\nu{}^\mu {}_\rho v^\rho \,,
\end{split}
\end{equation}
which adds to the original connection $\gamma_\mu$ the pull-back
to the string of the spacetime connection $\Gamma_\mu$ compatible with
the spacetime metric $G_{\mu\nu}$.
In general, the new connection contains both a four-dimensional and an embedding
contributions,
$ D_\alpha = \partial_\alpha+\gamma_\alpha +\partial_\alpha X^\mu \Gamma_\mu$.
On an arbitrary spacetime tensor, $T=T(X)$ it is easy to see that
\begin{equation} \label{eq:pullback-connection2}
\begin{split}
 D_\alpha T^{\mu \cdots}{}_{\nu \cdots} &=
 \partial_\alpha X^\rho \nabla_\rho T^{\mu \cdots}{}_{\nu \cdots} \,,
\end{split}
\end{equation}
where $\nabla_\mu$ denotes the spacetime connection compatible with $G_{\mu\nu}$.
Notice that we use the same symbol for brane's $\nabla_\mu$
and spacetime's $\nabla_\alpha$ covariant
derivatives to avoid overburdening the notation,
the distinction should be made obvious
by the indices.
By construction, the covariant derivative $D_\mu$
is compatible with both metrics $G_{\mu\nu}$ and $g_{\alpha\beta}$,
and allows us to introduce a helpful new Laplacian,
$\square \equiv g^{\alpha\beta} D_\alpha \partial_\beta$,
on the brane \cite{Ketov:2000dy,Flore:2012ma}.

Now we concentrate on the search for
conformal invariant operators constructed
with derivatives of $X^\mu$ and curvature tensors of the brane's geometry.
Under conformal transformations we require
\begin{equation} \label{eq:conformal-transformations}
\begin{split}
 X^\mu \to X^{\prime\mu}\,,
 \qquad g_{\alpha\beta} \to g'_{\alpha\beta}=e^{2\sigma}g_{\alpha\beta}\,,
\end{split}
\end{equation}
for some arbitrary local scalar function $\sigma(x)$.
The embedding coordinates thus transform with zero conformal weight
on the brane -- they are dimensionless coordinates as they should --
and consequently spacetime tensors, such as
$T^{\mu \cdots}{}_{\nu \cdots}$, do not transform either.
As a consequence, we seek for conformally invariant operators
of the form $\partial^n X^m$ that
must be contracted with the already invariant spacetime tensors.
A naive analysis based on scale invariance suggests that,
for a four-dimensional brane, such operators
must contain at least four derivatives with respect to the coordinates
on the brane.

Each instance of $X^\mu$ must be paired
with at least one derivative $\partial_\alpha$
so that the resulting operators are covariant with respect to
both brane's and spacetimes's transformations. Furthermore, the operators
must be densities in order to be integrated in an action,
therefore they will all be proportional to $\sqrt{g}$.
Since there must be four derivatives
in each operator, we can classify the basis of possible operators
using the number of $X^\mu$ that explicitly appear outside spacetime tensor dependencies.
This number ranges from zero to four, and the resulting operators will have
an equal number of uncontracted
spacetime indices.
In this construction, we have to take into account that each curvature tensor
of the brane, e.g.\ the brane's Riemann tensor $R_{\alpha\beta\gamma\delta}$,
counts as two derivatives,
$R_{\alpha\beta\gamma\delta} \sim \partial^2$.
All brane's indices must be contracted with an appropriate number of
the other available tensors on the brane, which are the metric $g_{\alpha\beta}$ and the Levi-Civita tensor $\epsilon_{\alpha\beta\gamma\delta}$.
For determining the conformal invariants we have adopted the most general ansatze
built from very lengthy linear combinations of operators
and determined the relative coefficients by brute force.

There is only one conformally invariant operator with no spacetime indices,
corresponding to the square of the brane's Weyl tensor
\begin{equation} \label{eq:conformal-invariant-0d}
\begin{split}
 {\cal O}_0 &= \sqrt{g} \, C_{\mu\nu\rho\theta}C^{\mu\nu\rho\theta} =\sqrt{g} \, C^2\,. 
\end{split}
\end{equation}
Another operator that might come to mind is the four-dimensional Euler's density
$\sqrt{g} E_4$, and in fact it will play an important role later,
but it is not actually conformal invariant, so we leave it outside our basis for now.
Moving on to one uncontracted index, we find that there is another operator
\begin{equation} \label{eq:conformal-invariant-1d}
\begin{split}
 {\cal O}^\mu_1 &= \sqrt{g} \Bigl\{ \square^2 
 + 2 R^{\alpha\beta} D_\alpha \partial_\beta 
- \frac{2}{3} R \,\square  +\frac{1}{3} \nabla^\alpha R \,\partial_\alpha 
 \Bigr\}X^\mu\,,
 \\&
 \equiv \sqrt{g} \hat{\Delta}_4 X^\mu
\end{split}
\end{equation}
The above structure is not unexpected as the differential operator
$\hat{\Delta}_4$ acting on $X^\mu$
is conformally covariant and we shall return to this point later.

\begin{widetext}

Also the case with two uncontracted indices contains a single
conformally invariant operator, that is
\begin{equation} \label{eq:conformal-invariant-2d}
\begin{split}
 {\cal O}^{\mu\nu}_2 = \sqrt{g} \Bigl\{
  &
  -4 \partial_\alpha X^{(\mu} D^\alpha\square X^{\nu)}
  -2 D_\alpha\partial_\beta X^\mu D^\alpha\partial^\beta X^\nu
  -\square X^{\mu} \square X^{\nu}
  -4 {R}^{\alpha\beta} \partial_{\alpha}X^\mu\partial_{\beta}X^\nu
  +\frac{2}{3} R \, \partial_{\alpha}X^\mu\partial^{\alpha}X^\nu
 \Bigr\}\,,
\end{split}
\end{equation}
which too has some relation with the aforementioned differential operator
$\hat{\Delta}_4$
as we are going to see explicitly soon.
The operator turns out to be symmetric in the spacetime indices,
${\cal O}^{\mu\nu}_2={\cal O}^{(\mu\nu)}_2$
and its arbitrary normalization has been chosen for later convenience.
The situation becomes even more interesting when constructing operators
with three external spacetime indices.
We find three distinct conformally invariant combinations
\begin{equation} \label{eq:conformal-invariant-3d}
\begin{split}
 {\cal O}^{\mu\nu\rho}_{3a} = \sqrt{g} \Bigl\{ &
 2 g^{\beta\gamma}\partial^\alpha X^{(\mu} \partial_\beta X^\nu D_\alpha \partial_\gamma X^{\rho)}
 + g^{\alpha\beta}\partial_\alpha X^{(\mu} \partial_\beta X^\nu \square X^{\rho)}
 \Bigr\}\,,
 \\
 {\cal O}^{\mu\nu\rho}_{3b} =\sqrt{g} \Bigl\{ &
 2 \partial^\alpha X^\mu \partial^\beta X^{(\nu} D_\alpha \partial_\beta X^{\rho)}
 + \partial_\alpha X^{\nu} \partial^\alpha X^{\rho} \square X^\mu
 \Bigr\}\,,
 \\
 {\cal O}^{\mu\nu\rho}_{3c} =\sqrt{g} \Bigl\{ &
 \partial^\alpha X^\mu \partial^\beta X^{[\nu} D_\alpha \partial_\beta X^{\rho]}
 -\partial_\alpha X^\mu \partial^\alpha X^{[\nu} \square X^{\rho]}
 \Bigr\}\,.
\end{split}
\end{equation}
We arranged the combinations in such a way that their symmetries
are as manifest as possible. Namely, the first is fully symmetric
${\cal O}^{\mu\nu\rho}_{3a}={\cal O}^{(\mu\nu\rho)}_{3a}$,
the second is symmetric in the last pair of indices,
${\cal O}^{\mu\nu\rho}_{3b}={\cal O}^{\mu(\nu\rho)}_{3b}$,
while the last one is antisymmetric in the same pair,
${\cal O}^{\mu\nu\rho}_{3c}={\cal O}^{\mu[\nu\rho]}_{3c}$.

\end{widetext}

Finally, there are two conformal operators with four spacetime indices
\begin{equation} \label{eq:conformal-invariant-4d}
\begin{split}
 {\cal O}^{\mu\nu\rho\theta}_{4} &= \sqrt{g}\,
 \partial_\alpha X^\mu \partial^\alpha X^\nu \partial_\beta X^\rho \partial^\beta X^\theta\,,
 \\
 {\cal O}^{\mu\nu\rho\theta}_{wz} &= 
 \epsilon^{\alpha\beta\gamma\delta}
 \partial_\alpha X^\mu \partial_\beta X^\nu \partial_\gamma X^\rho \partial_\delta X^\theta\,.
\end{split}
\end{equation}
The first operator is manifestly symmetric under the exchange of two symmetric pairs
${\cal O}^{\mu\nu\rho\theta}_{4}={\cal O}^{((\mu\nu)(\rho\theta))}_{4}$,
while the second one is totally antisymmetric
${\cal O}^{\mu\nu\rho\theta}_{wz}={\cal O}^{[\mu\nu\rho\theta]}_{wz}$
and it is the unique structure that we can form using the Levi-Civita tensor.
Notice that the totally antisymmetric operator contains $\sqrt{g}$
through the Levi-Civita tensor with raised indices.

A general action for the conformally invariant brane must include all the above
operators contracted with spacetime tensors with the correct indices and symmetries.
The action has thus the form $S[X]= \sum_i {\cal G}_i \cdot {\cal O}_i$, where
${\cal G}_i={\cal G}_i(X)$ are spacetime tensor that act as couplings
with the appropriate number of indices, the sum ranges over all
the above operators denoted collectively as
${\cal O}_i={\cal O}_i(X,\partial X)$, and the dot-product includes
a summation over the spacetime indices as well as an integration
over the coordinates on the brane.
We are going to display and discuss the action functional $S=S[X]$
more carefully in the next section. A similar analysis
of the Polyakov's string operator can be found in \cite{Osborn:1988hd,Osborn:1989bu}.

Some of the above operators can be written as total derivatives,
which is a property that is going to be useful later on, when restricting
our attention to a closed brane. In particular the operators
with an odd number of indices are all total derivatives
\begin{equation} \label{eq:total-der-operators}
\begin{split}
 {\cal O}^{\mu}_{1} &= \sqrt{g}\, D^\alpha J_{1,\alpha}^\mu\,,
 \\
 {\cal O}^{\mu\nu\rho}_{3q} &= \sqrt{g}\, D^\alpha J_{3q,\alpha}^\mu
 \qquad (q=a,b,c)\,,
 \end{split}
\end{equation}
where we introduced the following ``currents'' for the one-indexed operator
\begin{equation} \label{eq:current-operators-d1}
\begin{split}
 J_{1,\alpha}^\mu &= D_\alpha \square X^\mu
 +2 R_\alpha{}^\beta \partial_\beta X^\mu -\frac{2}{3} R \, \partial_\alpha X^\mu
 \,,
 \end{split}
\end{equation}
and for the three-indexed ones
\begin{equation} \label{eq:current-operators-d3}
\begin{split}
 J_{3a,\alpha}^{\mu\nu\rho} &=
 2g^{\beta\gamma}\partial_\alpha X^{(\mu} \partial_\beta X^\nu \partial_\gamma X^{\rho)}
 \,,
 \\
 J_{3b,\alpha}^{\mu\nu\rho} &=
 \partial_\alpha X^\mu \partial_\beta X^\nu \partial^\beta X^\rho
 \,,
 \\
 J_{3c,\alpha}^{\mu\nu\rho} &=
  \partial^\beta X^\mu \partial_\alpha X^{[\nu} \partial_\beta X^{\rho]}
 \,.
 \end{split}
\end{equation}
Finally, we can write $J_{1,\alpha}^\mu$ as the divergence of a symmetric
two-tensor on the brane with one spacetime index,
$J_{1,\alpha}^\mu = D^\beta Z_{\alpha\beta}^\mu$, as
\begin{equation} \label{eq:Stensor-d1}
\begin{split}
 Z_{\alpha\beta}^\mu = \frac{1}{3}\Bigl\{ D_\alpha \partial_\beta 
 +  g_{\alpha\beta} \square  + 4 R_{\alpha\beta} 
 -2 g_{\alpha\beta} R \Bigr\}X^\mu
 \,,
 \end{split}
\end{equation}
where we used the fact that the coordinates are spacetime scalars,
$D_\alpha X^\mu =\partial_\alpha X^\mu$. Notice, however,
that the final expression loses the manifest covariance with respect to
spacetime transformations since the $X^\mu$ appear without a derivative.
This implies that $Z_{\alpha\beta}^\mu$ is covariant under reparametrizations
of the brane, but not under spacetime transformations.

\section{The higher derivative action}\label{sect:hdpa}

Using the new Laplacian and the operators discussed in the previous section,
we can construct a higher derivative generalization of
the Polyakov action as linear combination by contracting the operators
with appropriate spacetime tensors.
We begin by noticing that the action must contain a unique
symmetric spacetime two-tensor contracting ${\cal O}_2^{\mu\nu}$. It
is straightforward to observe that the four derivatives operator
${\cal O}_2^{\mu\nu}$ is a natural generalization to four dimensions
of the two derivatives operator $\partial_\alpha X^\mu\partial^\alpha X^\nu$ contracting the metric in the bosonic string \eqref{eq:polyakov2d}.
We are thus lead to identify the tensor coupling
of ${\cal O}_2^{\mu\nu}$ with the metric $G_{\mu\nu}$ of spacetime.
The other tensors are, in principle, arbitrary, so the action must contain
several additional spacetime tensors compatible with the symmetries induced
by conformal invariance.
The main result of the analysis leads to the full brane action
\begin{equation}\label{eq:polyakov4d-full}
\begin{split}
 S =&\frac{1}{(4\pi)^2 \alpha'} \int d^{4} x \sqrt{g}
 \Bigl\{ G_{\mu\nu} \Bigl[
-2 D_\alpha\partial_\beta X^\mu D^\alpha\partial^\beta X^\nu
\\&
  -2 \partial_\alpha X^\mu D^\alpha\square X^\nu
  -2 D^\alpha\square X^\mu \partial_\alpha X^\nu 
  -\square X^{\mu} \square X^{\nu}
  \\&
  -4 {R}^{\alpha\beta} \partial_{\alpha}X^\mu\partial_{\beta}X^\nu
  +\frac{2}{3} R \, \partial_{\alpha}X^\mu\partial^{\alpha}X^\nu
  \Bigr]
  \\&
  + {\cal S}_{\mu\nu\rho}\Bigl[
  \square X^\mu \partial_\alpha X^\nu \partial^\alpha X^\rho
  +2 D_\alpha \partial_\beta X^\mu \partial^\alpha X^\nu \partial^\beta X^\rho 
  \Bigr]
  \\&
  + {\cal D}_{\mu\nu\rho}\Bigl[
   \partial_\alpha X^\rho \partial^\alpha X^\nu \square X^\mu
  +2  \partial^\alpha X^\mu \partial^\beta X^\nu D_\alpha \partial_\beta X^\rho
  \Bigr]
  \\&
  + {\cal A}_{\mu\nu\rho}\Bigl[
   \partial_\alpha X^\mu \partial^\alpha X^\nu \square X^\rho
  - \partial^\alpha X^\mu \partial^\beta X^\nu D_\alpha \partial_\beta X^\rho
  \Bigr]
 \\&
 + \Phi_{c}(X) \,C^2
 +{\cal T}_{\mu\nu\rho\theta} \, \partial_{\alpha} X^{\mu} \partial^{\alpha} X^{\nu} \partial_{\beta} X^{\rho} \partial^{\beta} X^{\theta} 
 \Bigr\}\,,
\end{split}
\end{equation}
which is weighted by a dimension two coupling $\alpha'$, that plays an
analog role to the one of the bosonic string with the same name.
From the above action we have intentionally left out two tensors that,
to some extent, are decoupled from the others.
They are the linear term sourcing the operator
${\cal O}_1^\mu$ of \eqref{eq:conformal-invariant-1d},
which can be helpful for setting up higher loop computations \cite{Howe:1986vm}, and
${\cal O}^{\mu\nu\rho\theta}_{wz}$ of \eqref{eq:conformal-invariant-4d}
that we are going to discuss later.

We have introduced several tensor couplings in \eqref{eq:polyakov4d-full}.
The simplest one is $\Phi_{c}=\Phi_c(X)$, which
is an arbitrary scalar tensor on the bulk spacetime,
that could remind the reader of the dilaton field introduced
in \eqref{eq:polyakov2d-eff},
but actually couples to the conformally covariant
scalar $C^2=C_{\alpha\beta\gamma\delta}C^{\alpha\beta\gamma\delta}$.
The tensor ${\cal T}_{\mu\nu\rho\theta}$ is symmetric under two pairwise exchanges
and the exchange of two pairs of indices,
${\cal T}_{\mu\nu\rho\theta}={\cal T}_{(\mu\nu)\rho\theta}
={\cal T}_{\mu\nu(\rho\theta)}={\cal T}_{\rho\theta\mu\nu}$,
but is otherwise arbitrary. The tensors ${\cal S}_{\mu\nu\rho}$,
${\cal D}_{\mu\nu\rho}$ and ${\cal A}_{\mu\nu\rho}$
couple to the operators of \eqref{eq:conformal-invariant-3d} in order,
and thus have the following symmetries:
${\cal S}_{\mu\nu\rho}={\cal S}_{(\mu\nu\rho)}$,
${\cal D}_{\mu\nu\rho}={\cal D}_{\mu(\nu\rho)}$
and ${\cal A}_{\mu\nu\rho}= {\cal A}_{\mu[\nu\rho]}$.
For the contractions of the terms involving the three-tensors
we have explicitly used their symmetries to make the full form of $S$
more compact.
One could check explicitly even at this stage that
\eqref{eq:polyakov4d-full} is fully invariant under the conformal transformations
\eqref{eq:conformal-transformations}.
As far as we know, this action has never been studied.

The action \eqref{eq:polyakov4d-full} is a higher derivative
nonlinear sigma model. Higher derivative actions are known to ne nonunitary,
and consequently oftentimes avoided,
because they give rise to negative norm states in the spectrum known as ghosts \cite{Pais:1950za}.
Even so, it should be evident that \eqref{eq:polyakov4d-full}
is much richer and more complex
than its two dimensional counterpart.
Considerable simplifications occur if we 
allow for integration by parts in \eqref{eq:polyakov4d-full},
which would correspond to studying a closed, e.g.\ boundaryless, brane.
The first two lines of
the integrand of \eqref{eq:polyakov4d-full} can easily be manipulated in the form
\begin{equation}\label{eq:manipulation0}
\begin{split}
 G_{\mu\nu} \square X^\mu \square X^\nu + {\rm \, total \,\, derivative}\,,
\end{split}
\end{equation}
and are already normalized to produce a naive higher derivative generalization
of the string's Polyakov action \eqref{eq:polyakov2d}.
The operators contracting with the spacetime three-tensors can be written as total derivatives, as shown in \eqref{eq:current-operators-d3}, therefore we can integrate
them by parts and hide their contributions in a redefinition of
${\cal T}_{\mu\nu\rho\theta}$. To show this in practice, consider for example
the fully symmetric interaction with ${\cal S}_{\mu\nu\rho}$,
which can be written as
\begin{equation}\label{eq:manipulation1}
\begin{split}
 &{\cal S}_{\mu\nu\rho}D^\beta \Bigl(
  \partial_\beta X^\mu \partial_\alpha X^\nu \partial^\alpha X^\rho
  \Bigr)\,.
\end{split}
\end{equation}
Integrating by parts we find
\begin{equation}\label{eq:manipulation2}
\begin{split}
 &{\cal S}_{\mu\nu\rho}D^\beta \Bigl(
  \partial_\beta X^\mu \partial_\alpha X^\nu \partial^\alpha X^\rho
  \Bigr)
  \\
  &=
  -\nabla_\nu {\cal S}_{\mu\rho\theta} \,
  \partial_\alpha X^\mu \partial^\alpha X^\nu \partial_\beta X^\rho \partial^\beta X^\theta
  + {\rm \, total \,\, derivative}\,,
\end{split}
\end{equation}
and the remaining term is automatically symmetrized as
${\cal T}_{\mu\nu\rho\theta}$ because of the contractions.
As a consequence, with a
redefinition of the tensor ${\cal T}_{\mu\nu\rho\theta}$ as implied
by \eqref{eq:manipulation2}, given below in \eqref{eq:T-def},
it is possible to rewrite the fourth line of \eqref{eq:polyakov4d-full}
as a total derivative. 

\section{More on the closed brane}\label{sect:closed}

As a consequence of the above manipulations,
if we concentrate on a brane with no boundary, we can use the above steps
to recover the action presented in Eq.\ \eqref{eq:polyakov4d-intro} of the introduction.
We repeat it here for convenience
\begin{equation}\label{eq:polyakov4d}
\begin{split}
 S_{\rm bl} =&\frac{1}{(4\pi)^2 \alpha'} \int d^{4} x \sqrt{g}
 \Bigl\{ G_{\mu\nu} \square X^{\mu} \square X^{\nu} 
 \\&
 -2 \Bigl[ {R}^{\alpha\beta} -\frac{1}{3} R \,g^{\alpha\beta}\Bigr] G_{\mu\nu} \partial_{\alpha}X^\mu\partial_{\beta}X^\nu 
 \\&
 + \Phi_{c}(X) \,C^2
 +T_{\mu\nu\rho\theta}\,\partial_{\alpha} X^{\mu} \partial^{\alpha} X^{\nu} \partial_{\beta} X^{\rho} \partial^{\beta} X^{\theta}
 \Bigr\}\,,
\end{split}
\end{equation}
where the new spacetime tensor $T_{\mu\nu\rho\theta}$ has the same symmetries of
${\cal T}_{\mu\nu\rho\theta}$ and they are related by the transformation
\begin{equation}\label{eq:T-def}
\begin{split}
 T_{\mu\nu\rho\theta}=&{\cal T}_{\mu\nu\rho\theta}
 -\frac{3}{2}\Bigl[
 \nabla_{(\nu} {\cal S}_{\mu)\rho\theta}
 +\nabla_{(\rho} {\cal S}_{\theta)\mu\nu}
 \\& 
 +\nabla_{(\nu} {\cal D}_{\mu)\rho\theta}
 +\nabla_{(\rho} {\cal D}_{\theta)\mu\nu}
 \Bigr]- \frac{1}{4}\Bigl[
 \nabla_{\theta} {\cal A}_{(\mu\nu)\rho}
 \\&
 +\nabla_{\rho} {\cal A}_{(\mu\nu)\theta}
 +\nabla_{\mu} {\cal A}_{(\rho\theta)\nu}+\nabla_{\nu} {\cal A}_{(\rho\theta)\mu}
 \Bigr]\,.
\end{split}
\end{equation}

Interestingly, the action \eqref{eq:polyakov4d} of the closed brane makes
the Weyl symmetry more manifest, to some extent.
In fact, a simple way to convince oneself that \eqref{eq:polyakov4d} is conformally invariant begins by observing that the kernel of the propagation of the $X^\mu$ fields,
as induced by the first two lines of \eqref{eq:polyakov4d},
in flat embedding spacetime is the differential operator
\begin{equation}\label{eq:ftpr-op}
\begin{split}
 \Delta_4 &= (\nabla^2)^2
 +2 R^{\alpha\beta}\nabla_\alpha \partial_\beta
 - \frac{2}{3} R \, \nabla^\alpha \partial_\alpha
 +\frac{1}{3}\nabla^\alpha R \, \partial_\alpha
 \,,
\end{split}
\end{equation}
that is known as Fradkin-Tseytlin \cite{Fradkin:1982xc}, or Paneitz \cite{paneitz},
or Riegert \cite{Riegert:1984kt} (FTPR) operator,
as it was independently discovered
several times.
The FTPR operator replaces here the two dimensional Laplacian (of the string)
because it is a conformally covariant operator
\cite{Erdmenger:1997gy,Erdmenger:1997wy} that transforms
as $\Delta_4 \to e^{-4 \sigma} \Delta_4$ for
$g_{\alpha\beta}\to e^{2\sigma} g_{\alpha\beta}$ in four dimensions.
Similarly to the two dimensional case, $\Delta_4$ could be generalized
to a conformally covariant operator in a general brane dimension
at the price of introducing nonderivative curvature based interactions for $X^\mu$,
which are however not allowed for a Polyakov-like embedding theory.
From this we deduce the uniqueness of \eqref{eq:polyakov4d}
as four-derivatives generalization of \eqref{eq:polyakov2d}
for a closed brane, having also recalled that
the last line of \eqref{eq:polyakov4d} involving the tensor $T_{\mu\nu\rho\theta}$
is manifestly Weyl invariant.
As a matter of fact, we could have \emph{engineered} the action
\eqref{eq:polyakov4d} specifically 
to produce a conformal operator of the form $\Delta_4$ upon integration by parts,
which is actually the original way in which we came across this action \cite{romoli}.

The above discussion on conformal invariance should not mislead the reader
to think that \eqref{eq:polyakov4d} and \eqref{eq:polyakov4d-full} are
the generalizations of a free (quadratic) theory with higher derivatives.
In fact, the equations of motion are
\begin{equation}\label{eq:eom4d}
\begin{split}
 &\hat{\Delta}_4 X^\mu
 - {\cal R}^\mu{}_{\nu\rho\theta}
 \partial_\alpha X^\nu \partial^\alpha X^\rho \square X^\theta
 + \frac{1}{2}\nabla^\mu \Phi_c \, C^2
 \\&
 -4 T^\mu{}_{\nu\rho\theta}
  \partial^\alpha X^\rho \partial^\beta X^\theta D_\alpha \partial_\beta X^\nu
 -2 T^\mu{}_{\nu\rho\theta} 
  \partial^\alpha X^\rho \partial_\alpha X^\theta \square X^\nu
 \\&
 +\frac{1}{2} \nabla^\mu T_{\nu\rho\theta\sigma}
  \partial_\alpha X^\nu \partial^\alpha X^\rho
  \partial_\beta X^\theta \partial^\beta X^\sigma
 \\&
 -2 \nabla_\sigma T^{\mu}{}_{\nu\rho\theta}
  \partial_\alpha X^\sigma \partial^\alpha X^\nu
  \partial_\beta X^\nu \partial^\beta X^\theta
 = 0\,,
\end{split}
\end{equation}
where the operator $\hat{\Delta}_4$
has already appeared in \eqref{eq:conformal-invariant-1d}.
This differential operator has
exactly the same form as $\Delta_4$ given in \eqref{eq:ftpr-op},
but is nontrivial in the internal space spanned by spacetime indices
(from the point of view of the brane),
because all covariant derivatives are replaced with the ones
involving the pull-back of the spacetime connection, $\nabla_\mu \to D_\mu$.

In the equations of motion \eqref{eq:eom4d}, derivative interactions are hidden
in the form of the connection of $D_\mu$, but also show up explicitly through
the tensor $T_{\mu\nu\rho\theta}$ defined in \eqref{eq:T-def} and the components
${\cal R}_{\mu\nu\rho\theta}$ of the Riemann
curvature of spacetime, $\left[\nabla_\mu,\nabla_\nu\right]={\cal R}_{\mu\nu}{}^\cdot{}_\cdot $,
for which we used a calligraphic symbol to more clearly distinguish
it from the curvature tensors on the brane.

\section{Towards quantum conformal invariance}\label{sect:scale-conformal}

Now we want to discuss the status of conformal
invariance of \eqref{eq:polyakov4d-full}
and \eqref{eq:polyakov4d} at the quantum level.
This step should intuitively require an analysis similar to the one that has lead
to \eqref{eq:polyakov2d-eff} for the string. An anomaly for conformal symmetry that
generalizes \eqref{eq:trace2d} is expected on the basis of the fact that
scalar fields conformally coupled with the operator $\Delta_4$ are known
to produce one.
The anomaly should be naturally expressed in terms of the basis of the 
operators ${\cal O}_i$ given in section \ref{sect:conformal}, with the eventual inclusion of conformal symmetry breaking operators (like the dilaton $\Phi$ coupling
of the string), and the final result must be evaluated
on-shell with \eqref{eq:eom4d}.
The analysis is, however,
considerably more complicate for the action with four derivatives,
so we approach it in steps.

An action that generalizes \eqref{eq:polyakov4d-full}
by including terms of order $\alpha'$ should at least include four new symmetric
spacetime two-tensors to accommodate the possibility that radiative corrections
depart the relative coefficients of the first three lines
from the values constrained by conformal invariance as in \eqref{eq:polyakov4d-full},
as well as many more operators that break conformal invariance at the level of
the tensors with three indices.
The resulting action would also be a generalized nonlinear sigma model which,
besides being very complicate and requiring
a local renormalization group-like analysis \cite{Osborn:1991gm}
such as the one of \cite{Osborn:1987au}, has simply not yet been studied in the literature.

In order to slightly simplify the problem, we concentrate on the boundaryless brane
with action \eqref{eq:polyakov4d}, thus basically allowing for partial integration
and elimination of the boundary terms. In this setup,
we introduce a generalization of \eqref{eq:polyakov4d} that follows the
spirit of \eqref{eq:polyakov2d-eff} of including conformal breaking terms
with a power of $\alpha'$ 
\begin{equation}\label{eq:polyakov4d-eff}
\begin{split}
 S_{\rm eff} =&\frac{1}{(4\pi)^2 \alpha'} \int d^{4} x \sqrt{g}
 \Bigl\{ G_{\mu\nu} \square X^{\mu} \square X^{\nu} + \Phi_c \, C^2
 \\&
 -2 \Bigl[ {R}^{\alpha\beta} -\frac{1}{3} R \,g^{\alpha\beta}\Bigr] G_{\mu\nu} \partial_{\alpha}X^\mu\partial_{\beta}X^\nu
 \\&
 + \partial_{\alpha} X^{\mu} \partial^{\alpha} X^{\nu} \partial_{\beta} X^{\rho} \partial^{\beta} X^{\theta} T_{\mu\nu\rho\theta}
 \\&
 + \alpha' {R}^{\alpha\beta}\partial_{\alpha}X^\mu\partial_{\beta}X^\nu P_{\mu\nu}
 + \alpha' {R}\partial_{\alpha}X^\mu\partial^\alpha X^\nu Q_{\mu\nu}
 \\&
 + \alpha' D_{\alpha} \partial_{\beta} X^{\mu} \partial^{\alpha} X^{\nu} \partial^{\beta} X^{\rho} A_{\mu\nu\rho}
 +\alpha' \sum \Phi \cdot {\cal I}
 \Bigr\}\,.
\end{split}
\end{equation}
In the above action
we followed a notation similar to the flat space case
\cite{Buchbinder:1988ei,Buchbinder:1991jw}
and included
two new symmetric tensors $P_{\mu\nu}$ and $Q_{\mu\nu}$,
as well as a tensor $A_{\mu\nu\rho}$ that is invariant under
the exchange of the last two indices, $A_{\mu\nu\rho}=A_{\mu(\nu\rho)}$.
It should not be confused with the conformally invariant three tensors
of \eqref{eq:polyakov4d-full}.

The contributions from $P_{\mu\nu}$, $Q_{\mu\nu}$ and $A_{\mu\nu\rho}$
are all departures from the original conformal symmetry of \eqref{eq:polyakov4d}:
the symmetric two tensors account for relative departures from the structure of
the first two lines of \eqref{eq:polyakov4d} (so only two tensors are needed
for the closed brane),
while $A_{\mu\nu\rho}$
is not invariant because it contracts with only one of the tensor structures
appearing in the original action \eqref{eq:polyakov4d-full}.

In \eqref{eq:polyakov4d-eff}
we have also compactly denoted further dilaton-like
curvature interactions as 
\begin{equation}\label{eq:dilaton-terms-4d}
\begin{split}
 \sum \Phi \cdot {\cal I} &= \Phi_{a} \, E_4 + \Phi_{r} \, R^2
 + \Phi_{a'} \square R \,,
\end{split}
\end{equation}
which forms a basis of curvature square terms on the brane.
The tensor
$E_4=R^{\alpha\beta\gamma\delta}R_{\alpha\beta\gamma\delta}
-4R^{\alpha\beta}R_{\alpha\beta}+R^2$
denotes the four-dimensional Euler density on the brane
which integrates to a topological invariant.
The scalar tensors $\Phi_i$ for $i=c,a,a'$
are labelled according to their role in the ``traditional'' notation \cite{Cappelli:2001pz}
of the four-dimensional conformal anomaly \cite{Antoniadis:1991fa,Antoniadis:1992xu}.
The two most important contributions
are the $c$- and $a$-anomalies, where the latter generalizes
the anomaly of the string discussed in the previous section \cite{Osborn:1991gm}.
As expected, all deviations from the conformally invariant action \eqref{eq:polyakov4d}
are weighted by an additional power of $\alpha'$ in the effective action \eqref{eq:polyakov4d-eff}.

The leading radiative corrections of the quantization
of \eqref{eq:polyakov4d-eff}
must include
contributions from the quantization of the fields $X^\mu$
of the four-dimensional brane, as well as
contributions from the integration over the nonconformal degrees of freedom
of the metric $g_{\mu\nu}$. The latter are equivalent to the $bc$-ghosts
resulting in
the $-26$ contribution to the anomaly of the string.
The complete computation of the
radiative corrections is still surprisingly complicated, so we 
\emph{approximate} the result by assuming that the two contributions
are decoupled in the leading renormalization of the effective action.

We simplify the problem by temporarily neglecting the nonconformal modes
of $g_{\mu\nu}$. Essentially, we regard the action \eqref{eq:polyakov4d-eff}
as the one of a generalized higher derivative
nonlinear sigma model in curved space.
The renormalization of this sigma model has been already carried
over in flat space in the past with several different methods
and target spaces
\cite{Buchbinder:1988ei,Buchbinder:1991jw,Percacci:2009fh,Flore:2012ma,Hasenfratz:1988rf}.
For the purpose of this paper, we have generalized the computation to curved space
using dimensional regularization \cite{romoli}
and covariant heat kernel methods for higher derivatives operators
\cite{Barvinsky:1985an,Barth:1985sy,Lee:1987mm}.
In this way, it is possible to find beta functions
for all the tensors appearing in \eqref{eq:polyakov4d-eff},
some of which have been already given elsewhere \cite{Percacci:2009fh}.
Even with the simplification
of a nondynamical metric, these beta functions are unwieldy,
mostly because of the tensor $A_{\mu\nu\rho}$, which allows for many
independent contractions of its indices, yet their complete form is known
in flat space \cite{Buchbinder:1991jw}.

For this presentation, we take $A_{\mu\nu\rho}=0$ and discuss the consistency
of this choice later.
In this limit, the beta function of $G_{\mu\nu}$
is like the one found for the string's nonlinear sigma model modulo an overall coefficient
\begin{equation}\label{eq:beta-G}
\begin{split}
\beta_{G_{\mu\nu}}=\alpha' {\cal R}_{\mu\nu}\,,
\end{split}
\end{equation}
which hints to the fact that we are on the right track
to produce Einstein-like equations as consistency equations
for the background spacetime.

Denoting $\tilde{P}_{\mu\nu}=-2G_{\mu\nu}+\alpha' P_{\mu\nu}$
and $\tilde{Q}_{\mu\nu}=\frac{2}{3}G_{\mu\nu}+\alpha' Q_{\mu\nu}$
the tensor coupling combinations contracting
with the operators of the general form $R\partial^2 X^2$,
we compactly find the beta functions (independently of $A_{\mu\nu\rho}$)
\begin{equation}\label{eq:betas-P-Q}
\begin{split}
\frac{\beta_{\tilde{P}_{\mu\nu}}}{2\alpha^\prime} =&\frac{4}{3} T_{\mu\rho\nu}{}^\rho
 + \frac{2}{3} \tilde{P}^{\rho\theta} T_{\mu\rho\nu\theta}
 +2{\cal R}_{\rho(\mu} \tilde{P}_{\nu)} {}^\rho + 2\nabla^2 \tilde{P}_{\mu\nu}
 \,,\\
\frac{\beta_{\tilde{Q}_{\mu\nu}}}{2\alpha^\prime}  =& 
 \frac{2}{3}({\cal R}_{\mu\nu}-T_{\mu\rho\nu}{}^\rho-T_{\mu\nu\rho}{}^\rho)
 +2{\cal R}_{\rho{(\mu}} \tilde{Q}_{\nu)} {}^\rho\\
 & + 2\nabla^2 \tilde{Q}_{\mu\nu} - (\tilde{P}^{\rho\theta}+4\tilde{Q}^{\rho\theta})(R_{\mu\rho\nu\theta}-T_{\mu\nu\rho\theta})
 \\&
 +\frac{1}{3}(\tilde{P}^{\rho\theta}+6\tilde{Q}^{\rho\theta})T_{\mu\rho\nu\theta}\,.
\end{split}
\end{equation}
\normalsize It is straightforward, but still rather nontrivial,
to check that $\tilde{P}_{\mu\nu}=-2G_{\mu\nu}$
and $\tilde{Q}_{\mu\nu}=\frac{2}{3}G_{\mu\nu}$ are actually
fixed point solutions
of the renormalization group flow of these two couplings (using also the form
of $\beta_{G_{\mu\nu}}$). This confirms
both the precise structure of \eqref{eq:polyakov4d} on the basis of
quantum scale invariance, and the first two lines
of \eqref{eq:polyakov4d-eff} \emph{independently} of the other tensors appearing
at the subleading order in $\alpha'$.

The real difficulties emerge when considering the beta functions
of the tensors $A_{\mu\nu\rho}$ and $T_{\mu\nu\rho\theta}$.
We find that a running for $A_{\mu\nu\rho}$ is generated by radiative corrections
\begin{equation}\label{eq:beta-A}
\begin{split}
 \beta_{A_{\mu\nu\rho}}
 &= \nabla_{(\nu} {\cal R}_{\rho)\mu}-2 \nabla_\mu {\cal R}_{\nu\rho} \,,
\end{split}
\end{equation}
and a similar relation holds for the tensor $T_{\mu\nu\rho\theta}$
\cite{Buchbinder:1988ei,Buchbinder:1991jw,Percacci:2009fh}.
In other words, neither the condition $A=T=0$, nor simply $A=0$,
are fixed points of the renormalization group
and their contributions to the final anomaly must be accounted for.\footnote{%
The possibility that the geometric condition on ${\cal R_{\mu\nu}}$ that leads to
$\beta_{A_{\mu\nu\rho}}=0$ in \eqref{eq:beta-A} has a geometrical meaning
has been indulged upon in \cite{Buchbinder:1988ei,Buchbinder:1991jw}.
However, in anticipation of the following,
we believe that the quantum $\beta_{A_{\mu\nu\rho}}$ contribution to the anomaly
cancels out exactly with a classical term with the same structure.
}
In fact, this highlights that the conformal anomaly
of \eqref{eq:polyakov4d-eff} must include tensor structures such as
those of
$A_{\mu\nu\rho}$ and $T_{\mu\nu\rho\theta}$ in \eqref{eq:polyakov4d-eff},
besides the usual ones associated to the $c$- and $a$-anomalies.
More generally, we stress again that the anomaly should also contain
all the tensor structures appearing in
the full action \eqref{eq:polyakov4d-full}.
From a different, but complementary, point of view,
the renormalization group of a generalized nonlinear sigma model
is defined up to a diffeomorphism of the emebedding space,
which implies that the anomaly accounts for both classical contributions
coming from \eqref{eq:polyakov4d-eff} and quantum contributions coming from
\eqref{eq:beta-A} and the other beta functions that must be evaluated \emph{on-shell}
using \eqref{eq:eom4d}.

For all the above reasons, we do not give further beta functions as they are not particularly illuminating, but rather try to directly give an estimate of the conformal anomaly. The complete computation of the anomaly is rather complicated, since it would
require the knowledge of the complete renormalization group behavior
of the open brane's action \eqref{eq:polyakov4d-full}, as well as that of the
$\alpha'$ corrections (some of which have appeared in \eqref{eq:polyakov4d-eff}).
All the computations should also take into account the integration in the path-integral of the conformal degrees of freedom of the brane's metric $g_{\mu\nu}$,
which is not known in general even for the simpler case of a quadratic scalar field
conformally coupled with the operator \eqref{eq:ftpr-op}.

To provide the estimate of the anomaly,
we include in it the contribution from
conformal higher derivative gravity
\cite{Fradkin:1981iu,Fradkin:1981hx},
which should be an appropriate approximation for the conformal degrees of freedom of $g_{\mu\nu}$, though it should be clear that we are missing the off-diagonal contributions coming from the mixing of the metric's and
coordinates fields.
We incorporate the renormalization group behavior of the dilatonic term $\Phi_a$, that corresponds to the so-called $a$-anomaly, because in four dimensions it plays a role analog to the string's $\Phi$,
and is found in \cite{deBerredo-Peixoto:2003jda}.
For simplicity, we assume that the known ambiguity of the $a'$ anomaly is resolved through the condition occasionally stated as ``$a=a'$''
\cite{Anselmi:1999bu,Anselmi:1999ut},
which is achieved by combining
the scalar fields $\Phi_a$ and $\Phi_{a'}$ in such a way that they couple to the brane's curvature $Q_4=E_4-\frac{2}{3}\square R$ \cite{Riegert:1984kt,Asorey:2003uf}. This combination of curvatures is chosen
because it has a convenient transformation under Weyl rescalings $\sqrt{g}Q_4 \to \sqrt{g}Q_4 \to \sqrt{g}(Q_4 + \Delta_4 \sigma)$, which is analog in form the transformation of $\sqrt{g}R$ for the two dimensional case.\footnote{%
The pertinent curvature that should be considered here is known as four-dimensional $Q$-curvature and is a geometric entity obtained as the nonderivative part of
a conformally covariant differential operator that generalizes $\Delta_4$ of \eqref{eq:ftpr-op} \cite{paneitz,branson}. We do not use the standard form of the $Q$-curvature here
because it would require a further redefinition of $\Phi_c$ too.
}

Finally, we go on-shell with \eqref{eq:eom4d} and
approximate the final result
by neglecting all possible anomalies besides
${\cal O}_2^{\mu\nu}$, $C^2$ and $E_4$.
We obtain
\begin{equation} \label{eq:trace4d}
 \begin{split}
  8\pi^2 \langle \Theta^\mu{}_\mu \rangle &={\cal O}^{\mu\nu}_2 B_{\mu\nu}^{(G)} + E_4 B^{(\Phi_a)} 
 + C^2 B^{(\Phi_c)}  \,,
 \end{split}
\end{equation}
where we defined the anomaly coefficients that contain both classical and quantum parts
\begin{equation} \label{eq:anomaly-coeffs}
 \begin{split}
  B_{\mu\nu}^{(G)} &=
  \alpha^\prime \Bigl({\cal R}_{\mu\nu} - 2\nabla_\mu \partial_\nu \Phi_a \Bigr)
  \,,
  \\ 
  B^{(\Phi_a)} &=
  \frac{7D}{90} -  \frac{87}{20} - \alpha^\prime \Bigl( \frac{1}{2}\nabla^2 \Phi_a + \partial_\mu \Phi_a \partial^\mu \Phi_a\Bigr)
  \,,
  \\ 
  B^{(\Phi_c)} &= \alpha^\prime \Bigl( \frac{199}{30} - \frac{D}{15}\Bigr)\,,
 \end{split}
\end{equation}
and we observe consistency with the case of a single complex scalar field shown in \cite{Fradkin:1981jc} (see also \cite{Fradkin:1983tg,Asorey:2003uf}).

We stress once more that in the above estimate of $\langle \Theta^\mu{}_\mu \rangle$ we have omitted all the operators ${\cal O}$, besides the one associated with the metric and the scalar ones associated with the famous $a$- and $c$-anomalies.
Nevertheless,
it might be possible
that all the anomaly terms that we do not display can be eliminated by
appropriately choosing the calligraphic tensors shown in \eqref{eq:polyakov4d-full}
as functions of the spacetime curvature ${\cal R}_{\mu\nu}{}^\rho{}_\theta$ and 
its covariant derivatives.
In fact, preliminary computations
in this direction, that we have already performed, agree with this statement;
for example, we see explicit cancellations of the $A_{\mu\nu\rho}$ anomaly
in special limits.
The reason why we cannot make a stronger statement is
because we still do not have access to the full computation,
but are limited to the boundaryless effective action \eqref{eq:polyakov4d-eff},
which implies that some relative
contributions to the anomaly are ``hidden'' by relations such as \eqref{eq:T-def}.
From now on we work on the assumption that it is possible to cancel the conformal anomaly.

All anomalies displayed in \eqref{eq:anomaly-coeffs} are expected to receive further contributions of order $\alpha'$ at two loops, just like in the case of the string.
This happens because the loop expansion and the $\alpha'$ expansion do not exactly match.
This is particularly relevant for the $c$-anomaly, which starts at order $\alpha'$
but is clearly missing further terms in order to set it to zero.
In fact, at this stage it is not clear if
it will be possible to set $B^{(\Phi_c)}=0$ in general,
but we now assume that it will.
On a slightly more positive note, we see that setting to zero the leading order of the $a$-anomaly lets us deduce a critical dimension for spacetime
\begin{equation} \label{eq:Dcrit-brane}
 \begin{split}
  D_{\rm cr} &= \frac{783}{14} \simeq 55.9 \,,
 \end{split}
\end{equation}
which disappointingly is not an integer. 
Even so, this estimate is incomplete
since, as previously stated,
we have neglected the
off-diagonal contributions between the nonconformal modes of the brane metric
and the fluctuations of $X^\mu$ when computing the functional trace.
Nevertheless, we assume that this value of $D_{\rm cr}$ is close to the true one 
and there is the possibility that the full computation will reveal an integer.\footnote{%
To very partially substantiate this possibility, notice first that in \eqref{eq:anomaly-coeffs} we have given the estimate of the anomaly contribution using conformal higher derivative gravity, which is based on the analogy that the $-26$ contribution to the string's anomaly in \eqref{eq:dilaton2d-coeff} comes from integrating the nonconformal degrees of freedom of the $2d$ metric. Alternatively, we could interpret the string's anomaly as $-26=-25-1$, where $-25$ is the full $2d$ gravitational contribution \cite{Aida:1996zn} and $-1$ is the subtracted $2d$ dilaton's contribution.
If we choose to estimate \eqref{eq:anomaly-coeffs} in analogy to the alternative interpretation, we should use the full higher derivative gravity's contributions of \cite{deBerredo-Peixoto:2004cjk}, which replaces $-\frac{87}{20}\to-\frac{196}{45}$ in $B^{(\Phi_a)}$ of \eqref{eq:anomaly-coeffs}, but also subtract the one of a $4d$ conformal scalar, which replaces $\frac{7D}{90} \to \frac{7(D-1)}{90}$ in $B^{(\Phi_a)}$. In this case the final result for the critical dimension is an integer, $D_{\rm cr}=57$, though we should deal with additional contributions in \eqref{eq:trace4d}.
The two results differ by the contributions of two standard scalar fields in $4d$ \cite{Salvio:2017qkx}.
}

The order $\alpha'$ the $a$- and the metric's anomalies can be understood 
as equations of motion of the low-energy effective action of the critical brane.
For diagrammatic and dimensional reasons, we believe that they are not going
to structurally change significantly with further and more complete computations.
We find that $B_{\mu\nu}^{(G)}=B^{(\Phi_a)}=0$ are equations of motion
of the low energy effective action
\begin{equation} \label{eq:brane-low-energy}
\begin{split}
 S &= \frac{1}{\kappa_0}\int d^D X \sqrt{G} e^{2 \Phi_a}
 \Bigl\{{\cal R}+4 G^{\mu\nu}\partial_\mu\Phi_a\partial_\nu\Phi_a
 \Bigr\} + {\cal O}(\alpha')
 \,,
\end{split}
\end{equation}
which in form is rather similar to \eqref{eq:string-low-energy},
but has the string's dilaton $\Phi$ replaced by its brane's analog $\Phi_a$.
There is also the significant difference that the sign
of the exponential factor that couples
the dilaton to the curvature term is opposite.
This implies that the Einstein frame
is obtained by defining the new metric
\begin{equation} \label{eq:Einstein-frame}
\begin{split}
 G'_{\mu\nu} = e^{\frac{4\Phi_a}{D-2}} G_{\mu\nu}
 \,,
\end{split}
\end{equation}
though, obviously, a large number of dimensions should be compactified
to obtain four-dimensional gravity.
Assuming that a generalization of the Curci-Paffuti relation
holds for the brane too \cite{Curci:1986hi}, or, more naively,
that $\Phi_a$ has a constant semiclassical expectation value,
we deduce that in the Einstein frame the gravitational coupling scales as
\begin{equation} 
\begin{split}
 \kappa^2 \sim \kappa_0^2 e^{-2 \langle \Phi_a \rangle} \,.
\end{split}
\end{equation}
Contrarily to the string's case,
the coupling decreases for increasing size of the dilaton.
Physical implications of the above results deserve to be discussed more
thoroughly in the future.

\section{Wess-Zumino term}\label{sect:further}

There is an important extension of \eqref{eq:polyakov4d-full}
that we want to discuss before moving on to the rest of this presentation.
When constructing \eqref{eq:polyakov4d-full}, we have omitted the inclusion
of the operator ${\cal O}_{wz}$ shown in \eqref{eq:conformal-invariant-4d}.
We can extend \eqref{eq:polyakov4d-full} through the inclusion of a topological
Wess-Zumino-Novikov-Witten term as
\begin{equation} \label{eq:WZNW4d}
\begin{split}
 \int d^4 x \,
 \epsilon^{\alpha\beta\gamma\delta} {\cal B}_{\mu\nu\rho\theta}\, 
 \partial_\alpha X^\mu \partial_\beta X^\nu
 \partial_\gamma X^\rho \partial_\delta X^\theta
 \,,
\end{split}
\end{equation}
where we introduced
the totally antisymmetric spacetime tensor
${\cal B}_{\mu\nu\rho\theta}$ generalizing
well-known sigma model results with special target spaces \cite{Witten:1983tw}.
It should be evident that the interaction involving
${\cal B}_{\mu\nu\rho\theta}$
is also a generalization to four dimensions of the one involving the Kalb-Ramond field $B_{\mu\nu}$, which couples to the two-dimensional bosonic string \eqref{eq:polyakov2d} as
\begin{equation} \label{eq:KB2d}
\begin{split}
 \int d^2 x \,
 \epsilon^{\alpha\beta} B_{\mu\nu}\, 
 \partial_\alpha X^\mu \partial_\beta X^\nu
 \,.
\end{split}
\end{equation}
The analogy between the two terms goes further if one
notices that ${\cal B}_{\mu\nu\rho\theta}$ are the components of a spacetime
$4$-form ${\cal B}$ and that its renormalization group beta function
is $\beta_{{\cal B}_{\mu\nu\rho\theta}}
= \alpha' \nabla^\lambda {\cal H}_{\lambda\mu\nu\rho\theta}$,
where
${\cal H}_{\lambda\mu\nu\rho\theta}= \nabla_{[\lambda} {\cal H}_{\mu\nu\rho\theta]}$
are the components of
the spacetime $5$-form ${\cal H}$
coming from covariant exterior differentiating ${\cal B}$
itself \cite{Buchbinder:1988ei,Buchbinder:1991jw}.
Among other things, the Kalb-Ramond field is known to provide a torsion
to the spacetime degrees of freedom. The four form ${\cal B}$ plays a similar
role in the case of the four-dimensional brane, leading to a natural generalization
of the connection \eqref{eq:pullback-connection}, which includes these effects
in the covariant expansion.

\section{Relation with $3$-brane}\label{sect:comparison}

Now it is important to discuss how the four-dimensional higher derivative brane
differs from and might relate to similar geometrical objects that have appeared in the String Theory literature before.
The natural comparison of the four brane is with the $p$-brane,
which is a $(p+1)$-dimensional object subject to Dirichlet boundary conditions
\cite{Johnson:2000ch}.
The physics of the $p$-brane is governed by a Nambu-Goto action of the form
\begin{equation} \label{eq:Dp-brane}
\begin{split}
 S_{p} &= M_{p} \int d^{p+1}x \sqrt{\det(\gamma_{\alpha\beta})}\,,
\end{split}
\end{equation}
where $\gamma_{\alpha\beta}= \partial_\alpha X^\mu \partial_\beta X^\nu G_{\mu\nu}$ is the pull-back of the metric $G_{\mu\nu}$ of the bulk spacetime under the embedding map with coordinates $X^\mu$. Of course, the pull-back is also the induced metric on the $(p+1)$-dimensional brane.
For simplicity and for consistency with the rest of the paper we are using the Euclidean formulation of the theory,
but it is trivial to switch to Minkowskian signature using the absolute value of the determinant of any metric. The constant $M_{p}$ has the dimension of a mass to the power $p+1$ and plays the role of tension for the brane.
Notice that \eqref{eq:Dp-brane} is the action of a $p$-brane, i.e.~a purely higher dimensional generalization of the Nambu-Goto action. It is \emph{not} the Dirac-Born-Infeld (DBI) action \cite{Leigh:1989jq},
which would be the proper brane's action that interacts with critical strings in the critical bulk dimensions.

Introducing the metric $g_{\alpha\beta}$ on the $p$-brane as an auxiliary field in \eqref{eq:Dp-brane},
we can recast the Nambu-Goto action in the Polyakov form
\begin{equation} \label{eq:Dp-brane-polyakov}
\begin{split}
 S_{p} &= \frac{M_{p}}{2} \int d^{p+1}x \sqrt{g} \Bigl\{
 G_{\mu\nu} \partial_\alpha X^\mu \partial^\alpha X^\nu - (p-1)
 \Bigr\}\,,
\end{split}
\end{equation}
from which it is evident that the limit $p=1$
coincides with the case of the string shown in \eqref{eq:polyakov2d},
allowing for the identification of $\alpha'$.
If we neglect the dynamics, the case $p=3$ is a four-dimensional object
that can naturally be compared with our four-dimensional brane \eqref{eq:polyakov4d}, in the sense that they are both classically described by the image of four-dimensional embedding maps $X^\mu$.

For our discussion we only need the simplest form of the action \eqref{eq:Dp-brane-polyakov} without any other assumption on symmetry, but it is worth to briefly mention some development spawning from the basic object.
The simplest generalization of the Polyakov action is the $11$-dimensional supermembrane theory, which is an anomaly-free source of $11$D supergravity,
though not renormalizable \cite{deWit:1988xki}.
The theory is known to have good quantum properties,
at least when concentrating on some of its sectors
\cite{Boulton:2002br}. Furthermore, recently, new ideas have emerged
on how to perturbatively quantize it
\cite{Lechtenfeld:2021zgd}. The core of these developments rests upon
the natural inclusion of supersymmetry.

Let us concentrate temporarily on the simplest case $p=1$ of \eqref{eq:Dp-brane-polyakov}. In this case, it is possible to quantize either action \eqref{eq:Dp-brane}
or \eqref{eq:Dp-brane-polyakov}: on the Nambu-Goto side there is the traditional
textbook procedure based on the light cone quantization over a bulk Minkowski spacetime, in which the critical dimension, previously seen in \eqref{eq:dilaton2d-coeff}, emerges as a condition to preserve Lorentz symmetry, that would otherwise be anomalous. On the Polyakov side we can follow the discussion summarized in Sect.~\ref{sect:2dpa}. The reason why the two procedures are consistent for $p=1$ is because
the Polyakov action \eqref{eq:Dp-brane-polyakov} is Weyl invariant, so the auxiliary field does not introduce new degrees of freedom, neither classically as in \eqref{eq:Dp-brane-polyakov}, nor quantum mechanically, given that we eliminate the conformal anomaly. A nonzero anomaly would otherwise provide a dynamics to the conformal mode of $g_{\alpha\beta}$.

Instead, if $p>1$, there is no consistent quantization of either \eqref{eq:Dp-brane} or \eqref{eq:Dp-brane-polyakov} 
as long as we do not invoke any further requirement like supersymmetry. On the Nambu-Goto side, one sees that
further geometric terms based on the extrinsic properties of the brane, e.g.~the extrinsic curvatures, are generated quantum mechanically \cite{Akama:2014fra,Forini:2015mca,deLeonArdon:2020crs,Goon:2020myi}. On the Polyakov side, one finds that the metric introduces new degrees of freedom
that propagate and thus differentiate the two formulations. One intuitive discussion of this problem is given in \cite{Tong:2009np}, where it is noticed that
objects with $p>1$ can classically develop spike/tube-like singularities in their geometry
without a balancing energy cost in the action. This is of course related to the fact that the conformal factor is propagating, but not bounded in the action, which is a well-known and longstanding problem in quantum gravity in $d\geq 2$ \cite{Mottola:1995sj,Martini:2021slj}.
Quantum mechanically, however, the bosonic spectrum of the Polyakov action for $p$-branes has good quantum properties \cite{Martin:2007kp},
but they still must face the problem of canceling the vacuum energy,
for which the natural solution is the inclusion of supersymmetry.

For the above reasons,
especially when looking at the problem from the Polyakov side,
it may be worthwhile to try to construct a higher dimensional object enforcing the requirement of Weyl invariance from the very start, while otherwise relaxing the requirement that its geometrical origin is based on a Nambu-Goto-like description. In this framework, the analysis of conformally invariant terms
for a four-dimensional object is the one that we have given in Sect.~\ref{sect:conformal}, which naturally results in the actions \eqref{eq:polyakov4d-full} and \eqref{eq:polyakov4d} depending on the boundary conditions.

One important point to make is that our attempt to the quantization of the four-dimensional brane and to the cancellation of the conformal anomaly is still partial, due to the complexity of the problem. For all that we know at the moment, the four-dimensional brane could reveal itself as an anomaly free object
and thus describe a complete and consistent brane, but it could also be impossible to cancel the anomaly in general. If the latter were to be the case, then 
\eqref{eq:polyakov4d-full} and \eqref{eq:polyakov4d} could only be regarded as effective models, much like the $p$-brane \eqref{eq:Dp-brane-polyakov}.

Consistent or not, the four-dimensional brane \eqref{eq:polyakov4d-full} should still interpolate with \eqref{eq:Dp-brane-polyakov} in the infrared, according to standard renormalization group arguments for effective theories.
We perform this analysis in the limit of small $\alpha'$ for the action \eqref{eq:polyakov4d-full} following its relevant deformations, which would thus be the infrared limit \emph{as relative to} \eqref{eq:polyakov4d-full}, where the original conformal
invariance is necessarily broken by the relevant terms to the renormalization group (an analog discussion for the DBI brane would require a more in-depth discussion of the breaking of conformal invariance and the small tension limit for the latter).
To elaborate on this,
imagine that we deform the action \eqref{eq:polyakov4d-eff} by including the renormalization group relevant terms
\begin{equation}\label{eq:polyakov4d-eff-deformation}
\begin{split}
 S_{\rm rel} =&\frac{1}{(4\pi)^2 \alpha'} \int d^{4} x \sqrt{g}
 \Bigl\{ H_{\mu\nu} \partial_\alpha X^{\mu} \partial^\alpha X^{\nu} + \Phi \, R + \Lambda
 \Bigr\}\,.
\end{split}
\end{equation}
In the relevant deformation we have yet another symmetric tensor $H_{\mu\nu}$
and another ``dilaton'' $\Phi(X)$, both dimension two deformations, as well as a dimension four deformation $\Lambda(X)$. The latter is generally known as the tachyon operator in the case of the string. The relevant deformations are in form the same operators of the string's Polyakov effective action \eqref{eq:polyakov2d-eff} and the Polyakov's $p$-brane. We have computed the renormalization group flow of the operators in \eqref{eq:polyakov4d-eff-deformation}, which can be expressed in terms of gamma-function, $\beta_{\cal O}=\gamma_{\cal O}\cdot {\cal O}$ where ${\cal O}= \{ H_{\mu\nu},\Phi,\Lambda\}$. The complete relevant flow is
\begin{equation}\label{eq:rg-relevant-deformation}
\begin{split}
 \beta^{(H)}_{\mu\nu} =  \alpha^\prime&\Bigl\{
 \Bigl({T}^\rho{}_{(\mu\nu)}{}^\sigma
 + 2{T}_{\mu\nu}{}^{\rho\sigma}
 -2{\cal R}^\rho{}_{(\mu\nu)}{}^\sigma \Bigr)H_{\rho\sigma}
 \\&
 +\Bigl(
 H_{\rho(\mu}{\cal{R}}_{\nu)}{}^\mu
 -\frac{1}{2}\square H_{\mu\nu}
 \Bigr)\Bigr\}
 \,,\\
  \beta^{(\Phi)} =  \alpha^\prime & \Bigl\{-\frac{1}{6}H_\mu{}^\mu+\tilde{Q}_{\mu\nu} H^{\mu\nu} + \frac{1}{4}\tilde{P}_{\mu\nu} H^{\mu\nu}-\frac{1}{2}\square \Phi \Bigr\} \, , \\
  \beta^{(\Lambda)} = \alpha^\prime &\Bigl\{\frac{1}{2}H_{\mu\nu}H^{\mu\nu}-\frac{1}{2}\square \Lambda \Bigr\}\, ,
\end{split}
\end{equation}
where in $\beta^{(\Lambda)}$ we have included the term proportional to $H^2$
that appears at the next-to-leading order in the renormalization of the composite tachyon operator, simply because it was directly accessible through our computation. Notice that the flows of $H_{\mu\nu}$ and $\Phi$ mix under
renormalization at the leading linear order, because they have the same canonical dimension.

Being the new operators relevant, any small deformation is driven to increasing size toward the infrared, so $S_{\rm rel}$ will dominate over $S_{\rm eff}$
at some infrared scale under very general arguments. One quirk of the flow is that if the tensors $P_{\mu\nu}$ and $Q_{\mu\nu}$ of \eqref{eq:polyakov4d-eff} are zero,
then $\tilde{P}_{\mu\nu}=-2G_{\mu\nu}$
and $\tilde{Q}_{\mu\nu}=\frac{2}{3}G_{\mu\nu}$
are the conditions of conformal invariance (see also the comments in relation to Eq.~\eqref{eq:betas-P-Q}),
which implies that $\beta^{(\Phi)} \propto \square \Phi$, as one would expect
from a proper dilaton at this order.
Therefore,
$\beta_{\cal O}$ is diagonal at the leading order in the set of relevant operators and the gamma-functions can be read directly.

Ideally, at some infrared scale determined by the ratio of $\alpha'$
and the intrinsic scale of the dimension two tensor $H_{\mu\nu}$, the relevant contribution will dominate over the conformal action \eqref{eq:polyakov4d}
and $H_{\mu\nu}$ can play the role of
the new bulk metric.
Assume for simplicity that $H_{\mu\nu}$ is nondegenerate
and related by a rigid scale transformation to $G_{\mu\nu}$,
$H_{\mu\nu}= \lambda G_{\mu\nu}$. We know that $\lambda \sim \mu^2$, where $\mu$
is the renormalization group scale.
Neglecting the $T$ contribution, that we have not yet fixed on the basis of conformal invariance, we have
\begin{equation}\label{eq:rg-relevant-deformation}
\begin{split}
 \beta^{(H)}_{\mu\nu} & = 3 \frac{\alpha'}{\lambda} {\cal R}^H_{\mu\nu}\,,
\end{split}
\end{equation}
which suggests that in the infrared limit the action is effectively the one of a $D3$-brane with the new expansion parameter $\alpha'/\lambda$ (${\cal R}^H$ is the curvature tensor of the new bulk metric). The above approximations should be relaxed to have a clearer picture of the low energy limit of the four brane. In particular, the relation between the two metrics can generally be nonrigid
and $H_{\mu\nu}$ be degenerate, but also the tensor $T$ can play a role. However, an educated guess for the conformal fixed point of $T$ would be $T_{\mu\nu\alpha\beta}\sim {\cal R}_{\mu(\alpha\beta)\nu}$, based on dimensional analysis and symmetry, which would not alter the final conclusion.

\section{Conclusions}\label{sect:conclusions}

We have undertaken the first steps toward the construction and quantization
of a conformally invariant four-dimensional brane,
which is described as a higher derivative
generalization of the Polyakov's string. The classical action of such brane
couples to several spacetime tensors, which are much richer than
the usual three of the string.
Among these tensors there are, most importantly,
the metric of spacetime $G_{\mu\nu}$;
two dilatons, which are associated to the standard anomalies of four-dimensional conformal field theories;
three tensors with three indices of various symmetries;
and two tensors with four indices,
of which one generalizes the Kalb-Ramond field.

The four-dimensional brane is quite clearly much more complex than the string
because it is rich of structures. In this respect, our analysis barely tapped
at the surface of its geometric content.
The complexity of the brane is decreased, to some extent, if the brane is closed
and, thus, without a boundary, in which case many of the aforementioned tensors
decouple from the spectrum.
This hints at the fact that boundary conditions for the open case
can be significantly important when coupling the brane to lower dimensional objects
or including local degrees of freedom on the boundary.

In order to discuss the quantum status of conformal symmetry of the model,
we have discussed three natural first steps toward
the covariant quantization: the renormalization group scale invariance of the theory,
the expectation value of the conformal anomaly, and the resulting low-energy
effective action.
We have approached these steps with increasing number
of simplifications and we have outlined the next steps that should be taken
to circumvent such simplifications. We believe that the final conditions
that ensure the full cancellation of the anomaly
should include a precise determination
of most of the aforementioned spacetime tensors in terms of
the curvature of spacetime itself.
The curvatures might even account for extrinsic terms \cite{Caselle:2014eka}.
The final result hints at a critical dimension
for the brane, which, unfortunately, we cannot yet estimate precisely,
but that should be computable with modern covariant techniques,
and at a low-energy effective action that shares
many similarities with the one of the string.

In fact, our original motivation was precisely the one of assessing
the uniqueness of the Polyakov's string in producing a low-energy
effective action for spacetime which includes general relativity
and the metric $G_{\mu\nu}$. From this point of view,
it is fair to say that the string might not be unique,
but it is certainly the simpler object to study and develop
covariantly.

In regards to the fact that all the additional tensors, besides the spacetime
metric $G_{\mu\nu}$, that are present in the four-dimensional brane's action
should be determined as function of the spacetime curvature ${\cal R}_{\mu\nu}{}^\rho{}_\theta$ to eliminate the conformal anomaly,
we have some interesting speculation.
It might well be that, once all tensors present in \eqref{eq:polyakov4d-full}
are determined, the final low-energy effective action is not the Einsten-Hilbert action with a dilaton, as we presented in \eqref{eq:brane-low-energy},
but instead a higher derivative curvature action coupled to two dilatons,
similar in form to Stelle's gravity \cite{Stelle:1976gc}.
This possibility would institute an interesting parallellism
between the Polyakov's actions and the gravitational actions at low-energies,
since the standard string action produces the standard Einstein-Hilbert action at low-energies, while the higher derivative brane's action could produce
the higher derivative Stelle's one in the same limit.
Higher derivative gravity has attracted interest over the years
because it is perturbatively renormalizable in four dimensions
and it is asymptotically free in some couplings,
although nonunitary
\cite{Julve:1978xn,deBerredo-Peixoto:2004cjk}.
This possibility is still open to investigation.

An important concern of our higher derivative action, shared in fact by any higher derivative action, is that
it is perturbatively nonunitary because of the presence of
ghost states in the spectrum \cite{Pais:1950za}.
Several proposals have been made, both in the past \cite{Lee:1969fy} and more recently \cite{Anselmi:2017lia,Mannheim:2020ryw},
on which is the correct way to circumvent the negative norm states.
It is also not entirely clear what are the implications of a nonunitary
brane for the unitarity of the underlying ``brane'' field theory.
Assuming that the four-dimensional brane can be consistently quantized,
and that all our approximations in the computation of the anomaly can be relaxed,
it is entirely possible that the four-dimensional brane
may play a role in the grander arena of higher dimensional objects
for string theory and generalized Galileons \cite{Chatzistavrakidis:2016dnj}.
One possibility could be to use the dimensional
continuation of the four-dimensional brane in $d=4-\epsilon$ down to
$\epsilon=2$
to describe some sort of nonperturbative ``exotic'' string.
Another possibility, restricted to the case $d=4$, emerges from the steps
that we have undertaken to include the traditional operators
of the Polyakov's string and the tachyon operator
as renormalization group relevant deformations
of the brane in section \ref{sect:comparison} (see also
\cite{romoli,Percacci:2009fh}). Our arguments point at the conclusion
that relevant deformations drive the four brane to a large scale limit
in which it effectively behaves as a traditional $D3$-brane.

\smallskip

\paragraph*{Acknowledgments.} This work originates from the master thesis
of MR discussed at the University of Pisa in December of 2020.
MR is grateful to S.~Bolognesi and A.~Vichi for valuable comments as referees
of the thesis. OZ is grateful to I.~L.~Shapiro for correspondence on
related topics.
Finally, we are especially grateful to A.~Tseytlin
for correspondence and suggestions
which resulted in a more careful analysis of all the conformal operators.



\begin{thebibliography}{99}

\bibitem{Callan:1985ia}
C.~G.~Callan, Jr., E.~J.~Martinec, M.~J.~Perry and D.~Friedan,
Nucl. Phys. B \textbf{262} (1985), 593-609

\bibitem{Dai:1989ua}
J.~Dai, R.~G.~Leigh and J.~Polchinski,
Mod. Phys. Lett. A \textbf{4} (1989), 2073-2083

\bibitem{Polyakov:1981rd}
A.~M.~Polyakov,
Phys. Lett. B \textbf{103} (1981), 207-210

\bibitem{Buchbinder:1988ei}
I.~L.~Buchbinder and S.~V.~Ketov,
Theor. Math. Phys. \textbf{77} (1988), 1032-1038

\bibitem{Buchbinder:1991jw}
I.~L.~Buchbinder and S.~V.~Ketov,
Fortsch. Phys. \textbf{39} (1991), 1-20

\bibitem{Hasenfratz:1988rf}
P.~Hasenfratz,
Nucl. Phys. B \textbf{321} (1989), 139-162

\bibitem{Percacci:2009fh}
R.~Percacci and O.~Zanusso,
Phys. Rev. D \textbf{81} (2010), 065012
[arXiv:0910.0851 [hep-th]].

\bibitem{Polyakov:1987zb}
A.~M.~Polyakov,
Mod. Phys. Lett. A \textbf{2} (1987), 893

\bibitem{Fradkin:1984pq}
E.~S.~Fradkin and A.~A.~Tseytlin,
Phys. Lett. B \textbf{158} (1985), 316-322

\bibitem{Fradkin:1985ys}
E.~S.~Fradkin and A.~A.~Tseytlin,
Nucl. Phys. B \textbf{261} (1985), 1-27
[erratum: Nucl. Phys. B \textbf{269} (1986), 745-745]

\bibitem{Tseytlin:1986ws}
A.~A.~Tseytlin,
Nucl. Phys. B \textbf{294}, 383-411 (1987)

\bibitem{Aida:1996zn} 
  T.~Aida and Y.~Kitazawa,
  Nucl.\ Phys.\ B {\bf 491}, 427 (1997)
  [hep-th/9609077].

\bibitem{Bonezzi:2021mub}
R.~Bonezzi, T.~Codina and O.~Hohm,
[arXiv:2103.15931 [hep-th]].

\bibitem{Curci:1986hi}
G.~Curci and G.~Paffuti,
Nucl. Phys. B \textbf{286} (1987), 399-408

\bibitem{Ketov:2000dy}
S.~V.~Ketov,
``Quantum nonlinear sigma models: From quantum field theory to supersymmetry, conformal field theory, black holes and strings,''
Springer (2000)

\bibitem{Flore:2012ma}
R.~Flore, A.~Wipf and O.~Zanusso,
Phys. Rev. D \textbf{87} (2013) no.6, 065019
[arXiv:1207.4499 [hep-th]].

\bibitem{Osborn:1988hd}
H.~Osborn,
Nucl. Phys. B \textbf{308} (1988), 629-661

\bibitem{Osborn:1989bu}
H.~Osborn,
Annals Phys. \textbf{200} (1990), 1

\bibitem{Pais:1950za}
A.~Pais and G.~E.~Uhlenbeck,
Phys. Rev. \textbf{79}, 145-165 (1950)

\bibitem{Howe:1986vm}
P.~S.~Howe, G.~Papadopoulos and K.~S.~Stelle,
Nucl. Phys. B \textbf{296} (1988), 26-48

\bibitem{Fradkin:1982xc}
E.~S.~Fradkin and A.~A.~Tseytlin,
Phys. Lett. B \textbf{110}, 117-122 (1982)

\bibitem{paneitz}
S.~M.~Paneitz,
SIGMA Vol. 4 (2008)
[arXiv:0803.4331 [math.DG]].

\bibitem{Riegert:1984kt}
R.~J.~Riegert,
Phys. Lett. B \textbf{134} (1984), 56-60

\bibitem{Erdmenger:1997gy}
J.~Erdmenger,
Class. Quant. Grav. \textbf{14} (1997), 2061-2084
[arXiv:hep-th/9704108 [hep-th]].

\bibitem{Erdmenger:1997wy}
J.~Erdmenger and H.~Osborn,
Class. Quant. Grav. \textbf{15} (1998), 273-280
[arXiv:gr-qc/9708040 [gr-qc]].

\bibitem{romoli}
M.~Romoli, Master Thesis, Universit\`a di Pisa (2020)

\bibitem{Osborn:1991gm}
H.~Osborn,
Nucl. Phys. B \textbf{363} (1991), 486-526

\bibitem{Osborn:1987au}
H.~Osborn,
Nucl. Phys. B \textbf{294} (1987), 595-620

\bibitem{Cappelli:2001pz}
A.~Cappelli, R.~Guida and N.~Magnoli,
Nucl. Phys. B \textbf{618} (2001), 371-406
[arXiv:hep-th/0103237 [hep-th]].

\bibitem{Antoniadis:1991fa}
I.~Antoniadis and E.~Mottola,
Phys. Rev. D \textbf{45} (1992), 2013-2025

\bibitem{Antoniadis:1992xu}
I.~Antoniadis, P.~O.~Mazur and E.~Mottola,
Nucl. Phys. B \textbf{388} (1992), 627-647
[arXiv:hep-th/9205015 [hep-th]].

\bibitem{Barvinsky:1985an}
A.~O.~Barvinsky and G.~A.~Vilkovisky,
Phys. Rept. \textbf{119} (1985), 1-74

\bibitem{Barth:1985sy}
N.~H.~Barth,
J. Phys. A \textbf{20} (1987), 875

\bibitem{Lee:1987mm}
H.~W.~Lee, P.~Y.~Pac and H.~K.~Shin,
Phys. Rev. D \textbf{35} (1987), 2440-2447

\bibitem{Fradkin:1981iu}
E.~S.~Fradkin and A.~A.~Tseytlin,
Nucl. Phys. B \textbf{201} (1982), 469-491

\bibitem{Fradkin:1981hx}
E.~S.~Fradkin and A.~A.~Tseytlin,
Phys. Lett. B \textbf{104} (1981), 377-381

\bibitem{deBerredo-Peixoto:2003jda}
G.~de Berredo-Peixoto and I.~L.~Shapiro,
Phys. Rev. D \textbf{70} (2004), 044024
[arXiv:hep-th/0307030 [hep-th]].

\bibitem{deBerredo-Peixoto:2004cjk}
G.~de Berredo-Peixoto and I.~L.~Shapiro,
Phys. Rev. D \textbf{71}, 064005 (2005)
[arXiv:hep-th/0412249 [hep-th]].

\bibitem{Salvio:2017qkx}
A.~Salvio and A.~Strumia,
Eur. Phys. J. C \textbf{78}, no.2, 124 (2018)
[arXiv:1705.03896 [hep-th]].

\bibitem{Anselmi:1999bu}
D.~Anselmi,
Class. Quant. Grav. \textbf{17} (2000), 2847-2866
[arXiv:hep-th/9912122 [hep-th]].

\bibitem{Anselmi:1999ut}
D.~Anselmi,
Phys. Lett. B \textbf{476} (2000), 182-187
[arXiv:hep-th/9908014 [hep-th]].

\bibitem{Asorey:2003uf}
M.~Asorey, E.~V.~Gorbar and I.~L.~Shapiro,
Class. Quant. Grav. \textbf{21} (2003), 163-178
[arXiv:hep-th/0307187 [hep-th]].

\bibitem{branson}
T.~P.~Branson,
Mathematica Scandinavica, 57, 293–345.

\bibitem{Fradkin:1981jc}
E.~S.~Fradkin and A.~A.~Tseytlin,
Nucl.\ Phys.\ B {\bf 203}, 157 (1982).

\bibitem{Fradkin:1983tg}
E.~S.~Fradkin and A.~A.~Tseytlin,
Phys. Lett. B \textbf{134} (1984), 187

\bibitem{Witten:1983tw}
E.~Witten,
Nucl. Phys. B \textbf{223} (1983), 422-432


\bibitem{Johnson:2000ch}
C.~V.~Johnson,
[arXiv:hep-th/0007170 [hep-th]].

\bibitem{Leigh:1989jq}
R.~G.~Leigh,
Mod. Phys. Lett. A \textbf{4}, 2767 (1989)

\bibitem{deWit:1988xki}
B.~de Wit, M.~Luscher and H.~Nicolai,
Nucl. Phys. B \textbf{320}, 135-159 (1989)

\bibitem{Boulton:2002br}
L.~Boulton, M.~P.~Garcia del Moral and A.~Restuccia,
Nucl. Phys. B \textbf{671}, 343-358 (2003)
[arXiv:hep-th/0211047 [hep-th]].

\bibitem{Lechtenfeld:2021zgd}
O.~Lechtenfeld and H.~Nicolai,
JHEP \textbf{02}, 114 (2022)
[arXiv:2109.00346 [hep-th]].

\bibitem{Akama:2014fra}
K.~Akama and T.~Hattori,
[arXiv:1403.5633 [gr-qc]].

\bibitem{Forini:2015mca}
V.~Forini, V.~G.~M.~Puletti, L.~Griguolo, D.~Seminara and E.~Vescovi,
J. Phys. A \textbf{48}, no.47, 475401 (2015)
[arXiv:1507.01883 [hep-th]].

\bibitem{deLeonArdon:2020crs}
R.~de Le\'on Ard\'on,
Class. Quant. Grav. \textbf{37}, no.23, 237001 (2020)
[arXiv:2007.03591 [hep-th]].

\bibitem{Goon:2020myi}
G.~Goon, S.~Melville and J.~Noller,
JHEP \textbf{01}, 159 (2021)
[arXiv:2010.05913 [hep-th]].

\bibitem{Tong:2009np}
D.~Tong,
[arXiv:0908.0333 [hep-th]].

\bibitem{Mottola:1995sj}
E.~Mottola,
J. Math. Phys. \textbf{36} (1995), 2470-2511
[arXiv:hep-th/9502109 [hep-th]].

\bibitem{Martini:2021slj}
R.~Martini, A.~Ugolotti, F.~Del Porro and O.~Zanusso,
Eur. Phys. J. C \textbf{81} (2021) no.10, 916
[arXiv:2103.12421 [hep-th]].

\bibitem{Martin:2007kp}
I.~Martin, L.~Navarro, A.~J.~Perez and A.~Restuccia,
Nucl. Phys. B \textbf{794}, 538-551 (2008)
[arXiv:0705.3692 [hep-th]].

\bibitem{Caselle:2014eka}
M.~Caselle, M.~Panero, R.~Pellegrini and D.~Vadacchino,
JHEP \textbf{01} (2015), 105
[arXiv:1406.5127 [hep-lat]].

\bibitem{Stelle:1976gc}
K.~S.~Stelle,
Phys. Rev. D \textbf{16} (1977), 953-969

\bibitem{Julve:1978xn}
J.~Julve and M.~Tonin,
Nuovo Cim. B \textbf{46} (1978), 137-152

\bibitem{Lee:1969fy}
T.~D.~Lee and G.~C.~Wick,
Nucl. Phys. B \textbf{9}, 209-243 (1969)

\bibitem{Anselmi:2017lia}
D.~Anselmi and M.~Piva,
Phys. Rev. D \textbf{96}, no.4, 045009 (2017)
[arXiv:1703.05563 [hep-th]].

\bibitem{Mannheim:2020ryw}
P.~D.~Mannheim,
Int. J. Mod. Phys. D \textbf{29}, no.14, 2043009 (2020)
[arXiv:2004.00376 [hep-th]].

\bibitem{Chatzistavrakidis:2016dnj}
A.~Chatzistavrakidis, F.~S.~Khoo, D.~Roest and P.~Schupp,
JHEP \textbf{03} (2017), 070
[arXiv:1612.05991 [hep-th]].

  
\end{thebibliography}
\end{document}